\documentclass[aps,prl,reprint,groupedaddress]{revtex4-1}
\usepackage{graphicx, xcolor,soul}
\begin{document}


\title{Insulating state and giant non-local response in an InAs/GaSb quantum well in the quantum Hall regime}


\author{Fabrizio Nichele}
\email[]{fnichele@phys.ethz.ch}
\homepage[]{www.nanophys.ethz.ch}
\affiliation{Solid State Physics Laboratory, ETH Z\"{u}rich - 8093 Z\"{u}rich, Switzerland}

\author{Atindra Nath Pal}
\affiliation{Solid State Physics Laboratory, ETH Z\"{u}rich - 8093 Z\"{u}rich, Switzerland}

\author{Patrick Pietsch}
\affiliation{Solid State Physics Laboratory, ETH Z\"{u}rich - 8093 Z\"{u}rich, Switzerland}

\author{Thomas Ihn}
\affiliation{Solid State Physics Laboratory, ETH Z\"{u}rich - 8093 Z\"{u}rich, Switzerland}

\author{Klaus Ensslin}
\affiliation{Solid State Physics Laboratory, ETH Z\"{u}rich - 8093 Z\"{u}rich, Switzerland}

\author{Christophe Charpentier}
\affiliation{Solid State Physics Laboratory, ETH Z\"{u}rich - 8093 Z\"{u}rich, Switzerland}

\author{Werner Wegscheider}
\affiliation{Solid State Physics Laboratory, ETH Z\"{u}rich - 8093 Z\"{u}rich, Switzerland}


\date{\today}

\begin{abstract}
We present transport measurements performed in InAs/GaSb double quantum wells. At the electron-hole crossover tuned by a gate voltage, a strong increase in the longitudinal resistivity is observed with increasing perpendicular magnetic field. Concomitantly with a local resistance exceeding the resistance quantum by an order of magnitude, we find a pronounced non-local resistance signal of almost similar magnitude. The co-existence of these two effects is reconciled in a model of counter-propagating and dissipative quantum Hall edge channels providing backscattering, shorted by a residual bulk conductivity.
\end{abstract}


\maketitle


An InAs/GaSb double quantum well (QW) sandwiched between two AlSb barriers shows a peculiar band alignment \cite{Kroemer2004}. A QW for electrons in InAs and a QW for holes in GaSb coexist next to each other. If the QWs’ thicknesses are small enough, a hybridization gap is expected to open at the charge neutrality point (CNP) \cite{Naveh1995,Leon1999}. Depending on the QWs' thicknesses and on the perpendicular electric field, a rich phase diagram is predicted \cite{Liu2008}. It should be possible to electrically tune the sample from standard conducting phases to insulating, semimetallic or topological insulator phases. Recent work on InAs/GaSb QWs showed signatures of topological phases in micron-sized Hall bars at zero magnetic field \cite{Knez2011,Knez2012,Suzuki2013}, as expected for the quantum spin Hall insulator regime \cite{Konig2007}. Beyond the topological insulator properties, that manifest themselves, the fate of topological edge states at finite magnetic field has not been investigated so far. Similarly to other semi-metals like graphene \cite{Abanin2007,Checkelsky2008} or CdHgTe/HgTe quantum wells \cite{Gusev2010,Gusev2012}, electron and hole Landau levels (LLs) can coexist close to the CNP \cite{Nicholas2000,Takashina2003}. A detailed understanding of the expected hybridization of LLs \cite{Tsay1997} and its manifestation in a transport experiment is still missing. 

Here we present magnetotransport measurements performed on gated InAs/GaSb double QWs. At high magnetic fields, in the electron and hole regimes, we observe the formation of standard LLs. Close to the CNP a peculiar state forms in which electrical transport is governed by counter-propagating edge channels of highly dissipative nature. We investigate the transport properties in this regime using different measurement configurations, and as a function of magnetic field and temperature. 

The experiments were performed on two devices (named device A and device B) obtained from the same wafer as described in Ref. \onlinecite{Charpentier2013}. In Ref. \onlinecite{Knez2010} a nominally identical structure was used, and a hybridization gap of $3.6~\rm{meV}$ was reported.
Hall bar structures were fabricated by photolithography and argon plasma etching. Device A consisted of a single Hall bar with a width of $25~\rm{\mu m}$ and a separation between lateral arms of $50~\rm{\mu m}$. Device B consisted of two Hall bars in series, oriented perpendicularly to each other. Their width is $25~\rm{\mu m}$ and the lateral voltage probes have various separations, the shortest being $50~\rm{\mu m}$ . Device A was covered by a $200~\rm{nm}$ thick $\rm{Si_3N_4}$ insulating layer, device B by a $40~\rm{nm}$ thick $\rm{HfO_2}$ layer. On both samples a Ti/Au topgate was deposited in order to tune the charge density. Except for the different capacitance per unit area due to the different dielectrics, the two devices showed comparable behavior.

The experiments above a temperature of $1~\rm{K}$ were performed in a $^4\rm{He}$ system with a maximum magnetic field of $7~\rm{T}$. The experiments at lower temperature were conducted in a $^3\rm{He}/^4{He}$ dilution refrigerator with a base temperature of $70~\rm{mK}$ and a maximum magnetic field of $12~\rm{T}$. Electric measurements were performed by low-frequency lock-in techniques using constant biases smaller than $20~\rm{\mu V}$ to avoid sample heating.

\begin{figure}
\includegraphics[width=\columnwidth]{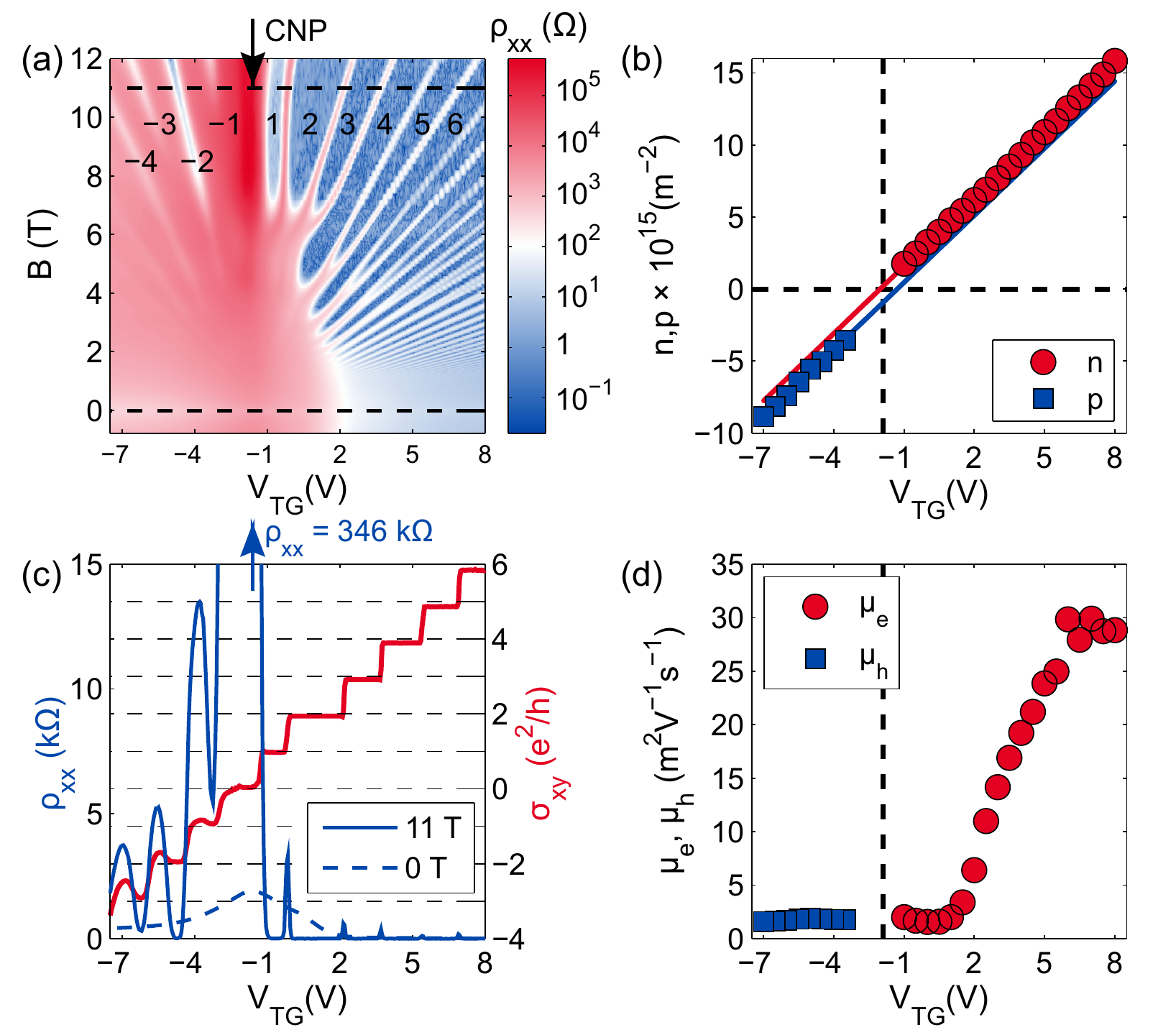}
\caption{(color online). (a) Longitudinal resistivity of sample A as a function of top gate voltage and magnetic field. The minima are labeled with the corresponding filling factor. The horizontal lines are cuts of the data visible in (c). (b) $n$ and $p$ as a function of top gate voltage, the lines are linear extrapolations of the data. (c) Resistivity at $11~\rm{T}$ and $0~\rm{T}$ (solid and dashed blue line respectively) together with the transverse conductivity at $11~\rm{T}$ (red line). (d) Carriers' mobility as a function of top gate voltage.}
\label{fig1}
\end{figure}

Fig. \ref{fig1}a shows the longitudinal resistivity of device A as a function of top gate voltage and perpendicular magnetic field measured at $70~\rm{mK}$. The device shows a pronounced ambipolar behavior where the occupation can be tuned from electrons in InAs to holes in GaSb (right and left side of Fig. \ref{fig1}a respectively). In the electron regime well developed SdH minima and spin-splitting are visible above $2~\rm{T}$. Oscillations in the hole regime can only be resolved at larger magnetic fields due to the higher effective mass. From the Hall slope it is possible to extract the electron and hole densities (${n}$ and ${p}$ respectively) far from the CNP. The results are shown in Fig. \ref{fig1}b. Both densities have a linear dependence on gate voltage with equal capacitances per unit area. Under the assumption of a constant density of states, a linear extrapolation of the data points indicates a partial band overlap and residual carrier densities of $\left(1.2\pm 0.09 \right)\times10^{14}~\rm{m^{-2}}$ at the CNP \cite{Pan2013} (see the lines in Fig. \ref{fig1}b). The energy overlap $\Delta$ between bands is equal to the Fermi energy when $p=0$, in our case we obtain $\Delta=6.4~\rm{meV}$. In Fig. \ref{fig1}d we show the carriers' mobility as a function of gate voltage. Close to the CNP, the mobility is about $2~\rm{m^{2}/Vs}$, and reaches $30~\rm{m^{2}/Vs}$ for high electron density. These value are comparable to those reported in Ref. \onlinecite{Knez2010}.

To estimate the disorder potential, we extracted the quantum scattering time $\tau_q$ from the temperature dependence of the low-field SdH oscillations in the electron regime. From $\tau_q=0.15~\rm{ps}$, we estimate the amplitude of the disorder potential to be $\hbar/\tau_q=6~\rm{meV}$, comparable to the band overlap. The large dimensions of the Hall bars and the large disorder potential do not allow the observation of the potential presence of helical edge states at zero magnetic field.

Fig. \ref{fig1}c shows two horizontal cuts of Fig. \ref{fig1}a for magnetic fields of $11~\rm{T}$ and $0~\rm{T}$ (solid and dashed blue line respectively) together with the transverse conductivity $\sigma_{xy}$ at $11~\rm{T}$ obtained by tensor inversion (red line). Similarly to Refs. \onlinecite{Knez2011,Knez2012}, the zero field resistivity shows a peak of the order of a few $k\Omega$ at the CNP. In a high magnetic field a prominent peak develops in $\rho_{xx}$ at the CNP and a $\rm{\nu}=0$ plateau appears in $\sigma_{xy}$. The well-developed minima in $\rho_{xx}$ and plateaus in $\sigma_{xy}$ at $11~\rm{T}$ away from the CNP indicate that the regime under study is governed by quantum Hall effect physics. The high resistivity peak ($346~\rm{k\Omega}$ at $11~\rm{T}$) is peculiar since it exceeds the resistance quantum by an order of magnitude.

\begin{figure}
\includegraphics[width=\columnwidth]{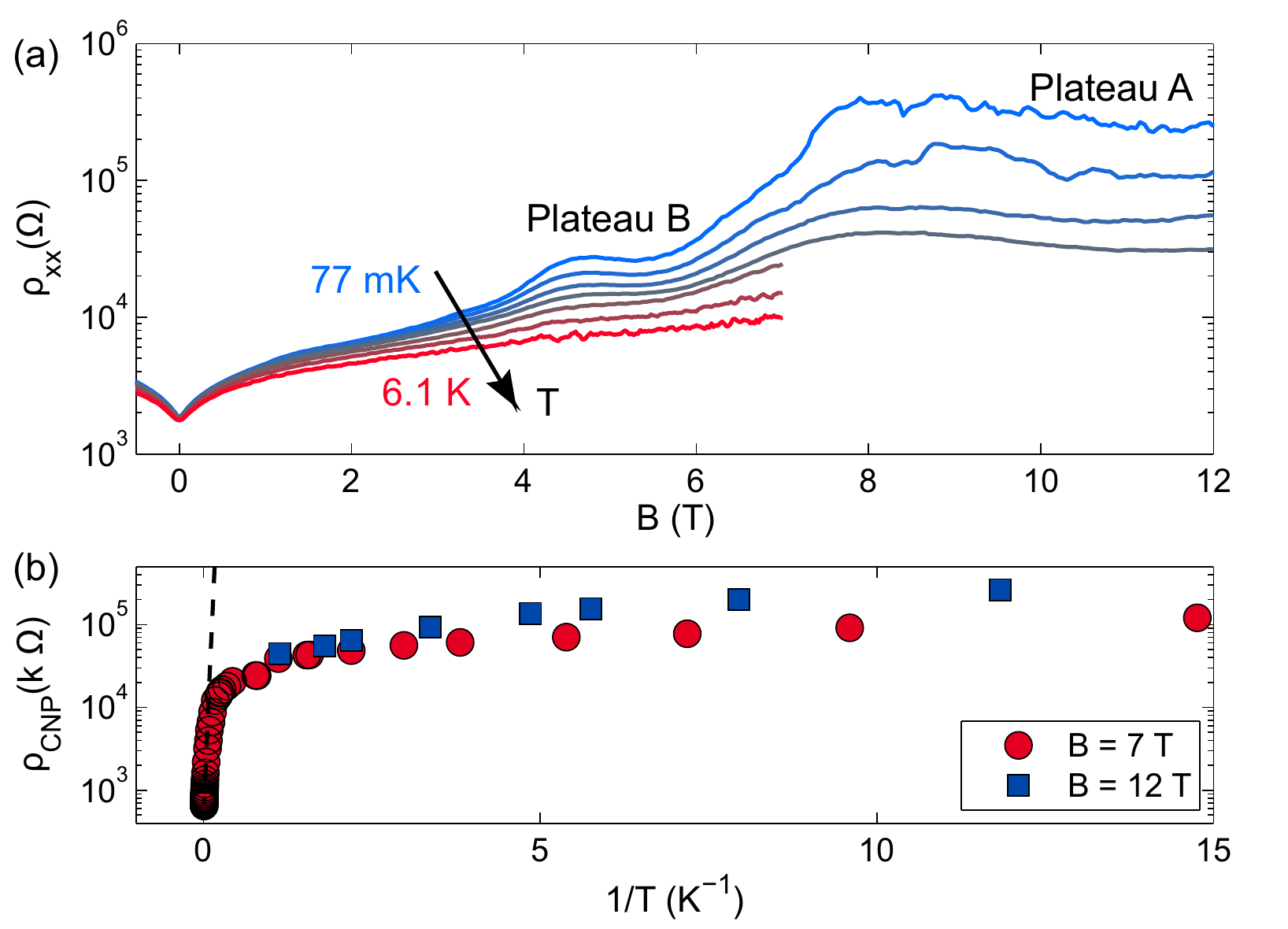}
\caption{(color online). (a) Temperature dependence of the resistivity at the CNP. (b) Arrhenius plot of the resistivity for different values of magnetic field.}
\label{fig2}
\end{figure}

Fig. \ref{fig2}a shows the longitudinal resistivity of device A at the CNP as a function of magnetic field. It increases from less than $2~\rm{k\Omega}$ at $\rm{B}=0~\rm{T}$ to $25~\rm{k\Omega}$ at $\rm{B}=4~\rm{T}$. For magnetic fields between $4~\rm{T}$ and $6~\rm{T}$, a plateau-like feature, labeled plateau B appears in $\rho_{xx}$ whose value at low temperature is close to $25~\rm{k\Omega}$. Above $6~\rm{T}$ the resistivity abruptly increases to a few hundred $\rm{k\Omega}$ (insulating state) and above $7.5~\rm{T}$ shows another plateau-like feature, labeled plateau A.

To check if the onset of the insulating state at the CNP is linked to the opening of an energy gap, we measured the temperature dependence of $\rho_{xx}$ at the CNP as a function of magnetic field. For small magnetic fields the temperature dependence is weak, and gets stronger as the magnetic field is increased. In Fig. \ref{fig2}b an Arrhenius plot of the CNP resistivity is shown for two different magnetic fields. The data at $7~\rm{T}$ has been measured up to $100~\rm{K}$, while the data at $12~\rm{T}$ only up to $900~\rm{mK}$. In the limit of high temperatures $\rho_{xx}$ strongly varies with temperature, indicating activated behavior while at low temperature the dependence is weak, which might indicate the onset of hopping transport. From the high-temperature slope (dashed line) an activation energy of $7.5~\rm{meV}$ was estimated. This value is of the same order of magnitude as the energy gaps between different LLs expected at $7~\rm{T}$ in this system. The logarithm of the low temperature resistivity can be fitted with a power law of the kind $T^{-1/\alpha}$, with $\alpha<4$. The limited data range did not allow to determine $\alpha$ with good precision. Hence the underlying hopping mechanism could not be identified.

\begin{figure}
\includegraphics[width=\columnwidth]{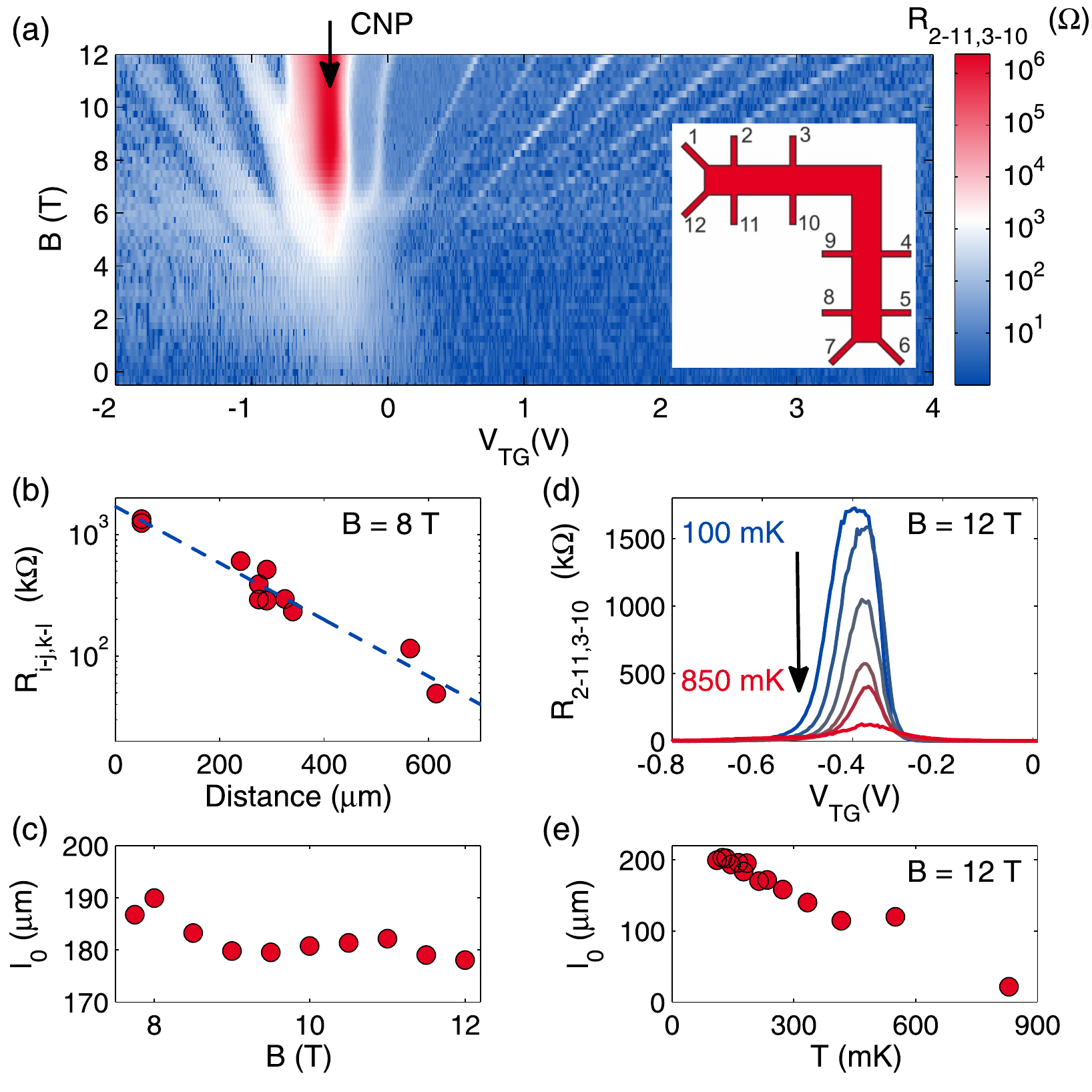}
\caption{(color online). (a) Contour plot of the four-terminal resistance $R_{2-11,3-10}$ as a function of magnetic field and top gate voltage for a $50~\rm{\mu m}$ contact separation. The inset shows a sketch of the device B. (b) Non local resistance at the CNP as a function of contact separation, the data is taken using various measurement configurations. The blue line is a fit to an exponential curve. (c) Decay length as a function of magnetic field. (d) Four-terminal resistance $R_{2-11,3-10}$ as a function of temperature at a constant magnetic field of $12~\rm{T}$. (e) Decay length as a function of temperature.}
\label{fig3}
\end{figure}

To confirm the presence of edge channels, we performed four-terminal measurements in device B using various contact configurations. A scheme (not to scale) of device B is shown in the inset of Fig. \ref{fig3}a. Such measurements are performed by passing a current $I_{i-j}$ between the pair of contacts $i$ and $j$, and measuring the voltage difference $V_{k-l}$ between the pair of contacts $k$ and $l$. The four terminal resistance is then defined as $R_{i-j,k-l}=V_{i-j}/I_{k-l}$. Fig. \ref{fig3}a shows a contour plot of $R_{2-11,3-10}$ as a function of magnetic field and gate voltage. In this case the voltage contacts are placed $50~\rm{\mu m}$ away from the direct path between the current contacts. At the CNP and at low magnetic field, the non-local resistance is smaller than $100~\Omega$. In correspondence to plateau A, a giant non-local response develops above $6~\rm{T}$ whose value of about $2~\rm{M\Omega}$ is much larger than the resistance quantum and comparable to the two-terminal resistance measured in the same configuration.

To study the dependence of the non-local response on the separation between current and voltage contacts, measurements have been performed with separations ranging from $50~\rm{\mu m}$ to $600~\rm{\mu m}$. Since device B has two Hall bars oriented perpendicularly to each other, we measure the distance between lateral arms along the central axis of the Hall bar. The results are shown in Fig. \ref{fig3}b for a magnetic field of $8~\rm{T}$. The signal decays as a function of length with a behavior that could be fitted with the exponential law (blue dashed line):
\begin{equation}
\label{eq1}
R_{i-j,k-l}(x)=R_{0}e^{-x/l_0},
\end{equation}
where $R_0$ is a constant prefactor and $l_0$ is the decay length. We could successfully fit the decay with Eq. \ref{eq1} for every magnetic field value above $7.75~\rm{T}$, obtaining values for $l_0$ close to $180~\rm{\mu m}$ (see Fig. \ref{fig3}c). This value of $l_0$ is surprisingly high. From standard diffusive transport a decay length of $W/\pi=8~\rm{\mu m}$ is expected \cite{Pauw1958}. This provides strong evidence for the presence of edge channel transport. Similar behavior was observed for any other measurement configuration in the L-shaped Hall bar, ruling out any transport anisotropy linked to the crystallographic orientation of the device.
On both sides of the CNP, we observe a fan of side peaks originating from the well-known non-local transport in the quantum Hall regime \cite{Takaoka1990, McEuen1990}. While the giant non-local peak at the CNP is always well visible, the side peaks can be distinguished from the noise just for the smallest distance between current and voltage probes.

Fig. \ref{fig3}d shows the temperature dependence of $R_{2-11,3-10}$ at $12~\rm{T}$. Similarly to $\rho_{xx}$, the non-local response is suppressed by temperature. The length $l_0$ decreases as well as the temperature is increased and, above a temperature of $1~\rm{K}$, a complete suppression of the non-local four-terminal resistance within the first $50~\rm{\mu m}$ is observed (see Fig. \ref{fig3}e).

\begin{figure}
\includegraphics[width=\columnwidth]{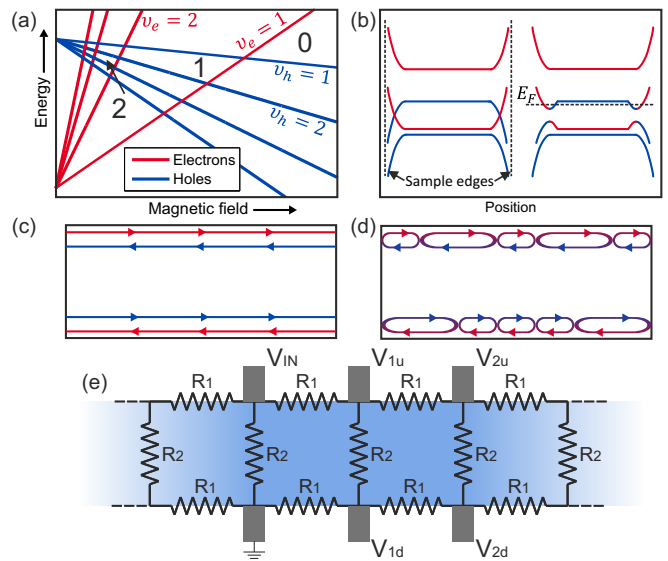}
\caption{(color online). (a) LLs for electrons (red) and holes (blue) without taking into account hybridization. (b) (Left) LLs' energy for electrons (red) and holes (blue) as a function of position along the sample. (Right) The same as on the left, but taking into account a hybridization between LLs. (c) ideal situation where two perfect counter-propagating edge channels are present. (d) Our case, where transport occurs along non-ideal edge channels. (e) The resistor model described in the text.}
\label{fig4}
\end{figure}

The effects described here can be understood in terms of hybridization of electron and hole LLs. If a band overlap is present, at high magnetic field LLs might hyridize \cite{Tsay1997, Vasilyev1999}. The expected LL spectrum (without hybridization) is sketched in Fig. \ref{fig4}a. In Fig. \ref{fig4}b (left) the first two electron and hole LLs are represented along a cross section of the Hall bar. The confinement potential at the sample edges pushes the LLs up or down depending on the charge sign and edge channels form. On the right of Fig. \ref{fig4}b the same situation is depicted including the hybridization of LLs. The hybridization creates a gap over the whole sample width. Shifting the Fermi energy through this gap we expect a transition from a situation where two counter-propagating edge channels are simultaneously present (as sketched in Fig. \ref{fig4}c) to one where transport is blocked. The latter case is never observed, since $\rho_{xx}$ always shows a measurable conductance and edge-channel transport is present. To account for the observed behavior the disorder potential has to be taken into account. A disorder potential comparable to the gap size can locally suppress the insulating state and give rise to carrier hopping between adjacent conducting puddles, leading to a finite resistance \cite{Nicholas2000,Takashina2003} (see Fig. \ref{fig4}d).

To study the different contributions in the resistance arising from the edges and bulk, we use an electrical model based on a resistor network as that depicted in Fig. \ref{fig4}c. The Hall bar is divided into finite elements, each having two local edge channel resistances $R_1$ and a local bulk resistance $R_2$. Applying a voltage $V_{in}$ to a contact with respect to a ground placed on the opposite side of the sample makes a current flow in the network. The four-terminal resistance, measured in the configuration described above, depends on $R_{1}$ and $R_{2}$. Considering the network to be of infinite length in both directions, it is found that $V_i$ scales exponentially with the element number $i$ such that a decay length $l_0$ and a prefactor $R_0$ can be defined (see Eq. \ref{eq1}). The calculation of the resistor model gives the following relations: ${R_1}/{R_2}=\mathrm{cosh}\left({W}/{l_0}\right)-1$ and ${R_0}=R_1\left(1+\sqrt{1+2R_1/R_2}\right)/2$.

Using $W=25~\rm{\mu m}$ and $l_0=180~\rm{\mu m}$ we obtain $R_1/R_2=1/100$ at base temperature, confirming that current preferentially flows along the edges.  These results confirm the applicability of the infinite chain model: since the Hall bar is much longer than $l_0$, all the current passes from one side to the other via the bulk, and not via edge channels. If no bulk conductance is allowed, the model should be modified into a finite chain of $R_1$ resistors and the voltages $V_i$ would then linearly decay over distance. Within our model, $l_{0}$ indicates the characteristic distance over which the edge channels on the two opposite edges of the sample equilibrate through bulk scattering.

Since the insulating state in $\rho_{xx}$ occurs between $\nu_e=1$ and $\nu_h=1$, we associate plateau A with the situation when $\nu_e=\nu_h=1$, as indicated in Fig. \ref{fig4}a by region 1. Plateau B can originate from the situation where $\nu_e=\nu_h=2$, as indicated by region 2. A very high magnetic field would bring the first electron LL above the first hole LL, as indicated by region 0, turning the sample into a normal band insulator. Our experimentally accessible magnetic field range does not allow probing this scenario. Temperature facilitates hopping transport along the edges and suppresses the localization in the bulk (both $R_{1}$ and $R_{2}$ decrease). $\rho_{xx}$ is mainly determined by the smaller, in our case $R_{1}$, but the non-local resistance depends on both $R_{1}$ and $R_{2}$ and is rapidly shunted by the onset of small bulk conductions.

We conclude that transport predominantly occurs at the sample edges and, since the two involved edge channels have different chirality, helical edge channels are expected. Checking the individual potential of every contact at high magnetic field, we found that the potential distribution along the edges follows opposite chirality for the electron and hole regime respectively. At the CNP no preferential direction is observed, confirming the helical nature of the edge channels under investigation. This situation is particularly interesting since it combines non-local transport, typically an effect arising from transport through ballistic edge channels, with a two terminal resistance exceeding by far the resistance quantum, which usually governs diffusive transport. Similarly to other experiments performed in GaAs \cite{Takaoka1990, McEuen1990, Geim1991}, the amplitude of the non-local response is determined by a competition between edge channels transmission and bulk conduction. The phenomena under consideration strongly differ from the ones known from standard electron transport. This is due to the different chirality and temperature dependent transmission of both bulk and edge channels.

In conclusion, we investigated the behavior of InAs/GaSb QWs at the CNP and at high perpendicular magnetic field. A strong resistivity increase accompanied by the onset of a giant non-local response was observed and studied. The results are interpreted in terms of helical quantum Hall edge states forming at high magnetic field.

\begin{acknowledgments}
The authors wish to thank L. Glazman, Y. Gefen and S. M\"{u}ller for constructive comments and useful discussions and the Swiss National Science Foundation for financial support via NCCR QSIT (Quantum Science and Technology).
\end{acknowledgments}

\bibliography{Bibliography}

\end{document}